\documentstyle[aps,prl,multicol]{revtex}
\begin{document}

\newcommand{\Dvec}{{\bf D}}
\newcommand{\Evec}{{\bf E}}
\newcommand{\Pvec}{{\bf P}}
\newcommand{\alp}{\alpha}
\newcommand{\epsi}{\epsilon}
\newcommand{\evac}{\varepsilon_{\rm v}}
\newcommand{\ez}{\varepsilon^0}
\newcommand{\einf}{\varepsilon^\infty}
\newcommand{\sia}{$\partial\sigma/\partial a$}
\newcommand{\sic}{$\partial\sigma/\partial c$}
\newcommand{\siu}{$\partial\sigma/\partial u$}
\newcommand{\kvec}{{\bf k}}
\newcommand{\vv}{\!\!\!\!\!\!}
\newcommand{\vare}{\varepsilon}  
\newcommand{\eps}{\epsilon}  
\newcommand{\De}{$\Delta$}
\newcommand{\de}{$\delta$}
\newcommand{\mcol}{\multicolumn}
\newcommand{\be}{\begin{eqnarray}}
\newcommand{\ee}{\end{eqnarray}}

\title{Polarization-based calculation of  the dielectric tensor of 
polar crystals}
\author{Fabio Bernardini and  Vincenzo Fiorentini}

\address{INFM -- Dipartimento di Scienze Fisiche, Universit\`a di 
Cagliari, I-09124 Cagliari, Italy}

\author{David Vanderbilt}
\address{Department of Physics and Astronomy, Rutgers University, 
Piscataway, NJ, U.S.A.}
\date{22 July  1997}
\maketitle

\begin{abstract}
We present a novel method for the calculation of the static 
and electronic dielectric tensor of polar insulating crystals
 based on concepts from the modern theory of   dielectric
polarization. As an application, we present the first ab initio
calculation of the dielectric constants in the wurtzite III-V nitrides
AlN, GaN, and InN. 
\end{abstract}

  
\begin{multicols}{2}
The modern quantum theory of polarization in dielectrics has been
formulated only in recent years \cite{polariz}. This
development has opened a new era in the first principles 
theory of  ferroelectricity and pyroelectricity 
\cite{ferro}. The new theory has been used successfully
to calculate, in a well defined
and computationally efficient way, the  macroscopic
polarization changes induced by perturbations other
than an electric field. Examples of such perturbations 
are e.g. lattice vibrations \cite{ferro},
ferroelectric distortions \cite{ferro}, and piezoelectric
 deformations \cite{piezo,piezo2}, whereby
the quantities being calculated are typically  dynamical Born charges,
spontaneous polarization, and piezoelectric constants.

So far, no direct attempt has been made  towards the goal 
of determining the  dielectric tensor (which of course quantifies 
the response to an external electric field) using polarization 
theory.  In this Letter we present a novel method for calculating 
the static dielectric tensor of a crystal based on concepts from
polarization theory; in particular, the method rests entirely 
on  the evaluation of the dielectric  polarization in zero field via
the  geometric quantum phase approach \cite{polariz}. 
 The method
works in any polar material, i.e. any material having infrared-active
zone-center modes.  As an
application, we provide the first (to our knowledge) determination of
the dielectric constants $\varepsilon_{\|}$ along the (0001) axis 
for the wurtzite III-V nitride compounds AlN, GaN, and InN.

The current method of choice for 
 dielectric response calculations  is Density Functional
Perturbation Theory (DFPT) \cite{DFPT},  a general and powerful
 approach to response properties. 
The  method presented here, 
besides its different foundations, is less general but
 considerably simpler to implement than DFPT,
and it may become a useful alternative.

\paragraph*{The static dielectric tensor -- }

The calculation of the dielectric tensor
 is highly non trivial, because it entails 
the determination of the electronic, as well as
vibrational  and elastic-piezoelectric, responses to an external 
electrostatic field. 
The elements of the dielectric tensor 
 are 
\be
\label{eq.ezero}
\ez_{ij}  = 
\delta_{ij}  + 4\pi \frac{d P_i}{d E_j}, 
\ee
where $\Evec$ is the screened macroscopic electrostatic field and 
$\Pvec$ the macroscopic polarization resulting from the 
 response of the electronic and ionic degrees of freedom.
In the Born-Oppenheimer approximation the macroscopic polarization 
$\Pvec$   in the presence of a generic strain or electrostatic  perturbation
can be conveniently expressed as 
$\Pvec = \Pvec^0 + \Pvec^{\rm lat}    + \Pvec^{\rm E }$,
the sum of
the spontaneous polarization $\Pvec^0$ of the
equilibrium structure in zero field, 
the polarization $\Pvec^{\rm lat}$ induced by lattice response,
and the electronic screening polarization $\Pvec^{\rm E}$.
  In the linear regime and using Voigt notation, the two
latter components  can be expressed in terms of the lattice structure
distortion and screened electric field  as
\be
\label{eq.pa}
P_i^{\rm lat} &=& \sum_{l} e^{(0)}_{il} ~\epsi_{l} 
	   +   \frac{e}{V} \sum_{sj} ~Z^{*\,s}_{ij} u^s_j\\ 
\label{eq.pe}
P_i^{\rm E}  &=& \frac{1}{4\pi} \sum_j  (\einf_{ij} - \delta_{ij}) ~E_j,
\ee
where
 $\einf_{ij}$
is the electronic component of the dielectric tensor,
 $V$ is the bulk cell volume, $\epsi_{l}$ the strain field,
$u^s_k$ the displacement of atom $s$ from its equilibrium
position; $e^{(0)}_{il}$ is the clamped-ion
component of the piezoelectric tensor, and
$Z^s_{ik}$  the Born effective charge:
\be
\label{eq.piezo}
\left. e^{(0)}_{il} = \frac{\partial P_i^{\rm lat}}{\partial \epsi_{l}}
                                                    \right\vert_u\, ,
\ \ \ \
\left. Z^{*\,s}_{ik} = \frac{V}{e} \frac{\partial P_i^{\rm lat}}{\partial
u^s_k}\right\vert
_\epsi \, .
\ee
The spontaneous component of the polarization does not depend 
on strain and fields, and it is non-vanishing
 also in zero field.
Using Eq. (\ref{eq.pa}) and (\ref{eq.pe}) we can rewrite Eq. (\ref{eq.ezero}) as
\be
\label{eq.pol}
  \ez_{ij} = \einf_{ij} + 
\left. 
   4 \pi \sum_l e^{(0)}_{il}
          \frac{\partial \epsi_{l}}{\partial E_{j}}
                                                  \right\vert_u
\left.
   +
   \frac{4 \pi e}{V} \sum_{sk}  Z^{*\,s}_{ik}
          \frac{\partial u^s_k}{\partial E_{j}}
                                                  \right\vert_\eps
\ee
We shall discuss first how the second and third term are evaluated, and 
 then discuss the electronic dielectric tensor.
The last two terms in Eq. (\ref{eq.pol}) quantify  
the contributions of the macroscopic 
and microscopic structural degrees of freedom  to the 
total polarization, respectively.
In the absence of other external perturbations, the
strain field $\epsi_{l}$ and the atomic displacement $u^s_k$
are related to the screened field $\Evec$
by the condition of vanishing stress,
\be
\label{eq.sigma}
 \sigma_{l} =  \sum_i e_{il} E_i
- \sum_{m} \lambda_{lm}  \epsi_{m} = 0 \, ,
\ee
and vanishing Hellmann-Feynman forces,
\be
\label{eq.force}
F^s_i =    e\, \sum_j  Z^{*\,s}_{ij} E_j
  + V\, \sum_{ls}  \Xi^s_{il} \,\epsi_l 
  - \sum_{js'} \Phi^{s\,s'}_{ij} u^{s'}_j  
  =  0 .
\ee
The quantities appearing in Eqs. (\ref{eq.sigma}) and
\ref{eq.force} are the 
elastic constants
\be
\label{eq.lambda}
   \lambda_{lm} = 
-  \frac{\partial \sigma_l}{\partial \epsi_m}
-  \sum_{is'} 
   \frac{\partial \sigma_l}{\partial u^s_i}
   \frac{d u^s_i}{d \epsi_m},
\ee
 the piezoelectric tensor
\be
\nonumber
e_{il}& =& \frac{\partial P_i^{\rm lat}}{\partial \epsi_{l}}
+  \sum_{js} 
   \frac{\partial P^{\rm lat}_i}{\partial u^s_j}
   \frac{d u^s_j}{d \epsi_l} \\ & =  &
e^{(0)}_{il} 
+  \sum_{jkss'}  Z^{*\,s}_{ij} \Phi^{-1\, ss'}_{jk} \Xi^{s'}_{kl},
\ee
and the harmonic force constants and  internal strain parameters
\be
\label{eq.phi}
  \Phi^{ss'}_{ij} = 
  \left.
 -\frac{\partial F^s_i}{\partial u^{s'}_j}\right\vert_\epsi,
\ \ \  \Xi^s_{il}  = \frac{1}{V} 
  \left.
  \frac{\partial F^s_i}{\partial \epsi_l}\right\vert_u.
\ee
Combining Eq.~(\ref{eq.sigma}) and~(\ref{eq.force}), we obtain 
 $u$ and $\epsi$ as a function of the  electric field: 
\be
u^s_i &=& \sum_{jks'} \Phi^{-1\, ss'}_{ik} \left( e Z^{*\,s'}_{kj} 
	     + V\, \sum_{lm} \Xi^{s'}_{kl} \lambda^{-1}_{lm} e_{mj} 
	     \right) E_j\, ;
	     \\
\epsi_i  &=& \sum_{jk} \lambda^{-1}_{ik} e_{kj} E_j\, .
\ee
These relations are the first two key ingredients of our method.
Substituting them into Eq.~(\ref{eq.pol}), we obtain after some manipulation
the following general expression for the static dielectric tensor:
\be \nonumber
  \ez_{ij} & = & \frac{4\pi e^2}{V}   
    \sum_{klss'} Z^{*\,s}_{ik}\, \Phi^{-1\,ss'}_{kl}\,
    Z^{*\,s'}_{lj}\\
&  + & 4\pi  \,\sum_{mn} e_{im}\, \lambda_{mn}^{-1}\, e_{nj} 
 + \einf_{ij} \label{eq.e0}  \\
\nonumber 
& = &  \vare^a +  \vare^b + \einf_{ij}. 
\ee
%
As implied by
Eq.~(\ref{eq.sigma}),  this definition holds for fixed stress, 
certainly an experimentally relevant situation.
The first term in the last equation
is due to atomic displacements
from the ideal position 
{ at fixed lattice parameter}; the second is 
the contribution of  piezoelectricity-related
lattice constant changes; the third  is the
pure electronic dielectric screening for a frozen lattice 
system.

A central point of the above analysis is that
all the ingredients of Eqs.\ (\ref{eq.piezo}) and 
(\ref{eq.lambda}$-$\ref{eq.phi}) needed for Eq.\ (\ref{eq.e0}) can 
 be calculated from distorted and strained bulk cells using ab-initio
total-energy and force calculations, supplemented by calculations of the
dielectric polarization in zero field using the Berry-phase
approach \cite{polariz,piezo,piezo2,nota}. The 
only exception is the electronic dielectric constant 
$\varepsilon^\infty$, 
for  which \cite{nota} an alternate approach has to be devised.

\paragraph*{The electronic dielectric tensor -- }
As we now show,  $\einf$ can be obtained using the relationship 
between macroscopic polarization in zero field and charge accumulation
at the interfaces of  an appropriately built  homojunction of the
material of interest \cite{noi.mrs,Vanderbilt.PRB48}.
In an insulating superlattice consisting of periodically alternating 
slabs of equal length, stacked along 
direction $\hat{\bf n}$ and made of materials 1 and 2,
the displacement field orthogonal to the interfaces is conserved:
$D_1 = E_1 + 4 \pi P_1 (E_1) =
 E_2 + 4 \pi P_2 (E_2) = D_2.$
(We use  a scalar notation for the components of the vectors along 
$\hat{\bf n}$.)
Expanding the polarization to first order in the screened fields in
the two materials as
$P_i (E)= P_i + \chi_i E_i$, with $P_i$ the polarization in zero
 field and $\chi_i$ the susceptibility, one obtains 
$$ 4 \pi (P_2 - P_1) = \varepsilon^{\infty}_1 E_1 -
\varepsilon^{\infty}_2 E_2.$$ 
 In the absence of zero-field (e.g. spontaneous) polarization, the
familiar equality  
$\varepsilon^{\infty}_1 E_1 =
\varepsilon^{\infty}_2 E_2$ 
is recovered.
To proceed, we note that 
periodic boundary conditions imply  $E \equiv E_1=-E_2$,
and $\Delta E \equiv E_1 - E_2 = 2 E$, so that 
$$ 4 \pi (P_2 - P_1) = \frac{1}{2}(\varepsilon^{\infty}_1 +
\varepsilon^{\infty}_2) \Delta E.$$ 
The charge accumulation per unit  area
at the interface between materials 1 and 2 is
$s_{\rm int} = \pm \Delta E / 4\pi$. Therefore,
switching to an obvious vector notation,
\be \label{eq.einf}
  s_{\rm int} = \pm 2\, {\bf \hat n}\cdot(\Pvec_2 - \Pvec_1) /
(\varepsilon^{\infty}_1 + \varepsilon^{\infty}_2 ),
\ee
which connects the 
 macroscopic bulk polarizations $\Pvec_{1,2}$  at zero field
with  the components
$\varepsilon^{\infty}_{1,2}$
 of the dielectric tensors of the interfaced materials
along the interface  normal ${\bf \hat n}$. 

In an undistorted
 homojunction, i.e. a superlattice in which material 1 is identical to 
material 2, there is no interface,
 no polarization change can occur, and  the  interface charge
is zero. However, a polarization difference 
can be  generated in a controlled manner by 
inducing  a small distortion $\delta$
of one of the atomic sublattices in half of the superlattice
unit cell. The interface  charge $s_{\rm int}$ accumulated 
at the interface  between distorted and undistorted
regions can   be easily  calculated via macroscopic averages
 \cite{noi.mrs,resta}. The  zero field 
polarizations $\Pvec_2$ for the material
in the undistorted state,
and $\Pvec_1$ for the material in the same strain state
as in the superlattice, are evaluated directly using
the Berry phase technique. From Eq. (\ref{eq.einf}),
one then  extracts the
 average electronic dielectric constant $\bar\vare^{\infty} = 
(\varepsilon^{\infty}_1 + \varepsilon^{\infty}_2 )$/2.  

In principle $\vare_1$, the dielectric constant in the 
distorted state,
differs from the actual
 dielectric constant  $\vare_2$; thus, so does $\bar\vare$.
However, in the limit of zero distortion, $\bar\vare$  equals the 
component of the dielectric tensor along $\hat{\bf n}$: $$
\varepsilon^{\infty} = \lim_{\delta\rightarrow 0}\, \bar\vare.$$
This limit can be evaluated  with essentially
arbitrary accuracy by extrapolation or interpolation.
The procedure just outlined, yielding the electronic dielectric 
constants, is the third key ingredient of the method.

\par
In summary,
in the present approach the static dielectric constant 
is calculated via  (i) 
calculation of the elastic and force constants, 
(ii) calculation of the piezoelectric 
tensor and Born charges, (iii) evaluation of the electronic 
dielectric tensor.
Task (i) requires standard  
total energy and stress calculations; tasks (ii) and (iii) 
 use the 
geometric quantum phase polarization; task (iii) also uses
 relatively small, accurately controllable supercell
calculations. In the latter, one must take care that {\it (a)} the
slabs be short  
enough that the constant electric field will not cause metallization,
and that {\it (b)} the slabs be sufficiently long to recover bulk-like
behavior away from the interfaces. Both requirements are generally met 
 even by materials with small calculated gaps for sufficiently
small applied  strains.

The quantities needed to evaluate the dielectric tensor are usually
 obtained by means of 
DFPT \cite{DFPT,resta2}.
The novelty of the present method  is in the absence of a perturbative 
approach, and in the determination of the  electronic screening and
 piezoelectric  properties using their connection with the geometric 
quantum phase.

\paragraph*{Application to III-V nitrides -- }
We now apply the formalism just developed 
to the calculation of $\ez_{33}\vert_a$, the 
component of the static dielectric tensor  along
the (0001) axis at fixed lattice constant $a$, for 
the wurtzite III-V nitrides AlN, GaN, and InN \cite{nota2}.
Besides serving as a test of our theory,
this calculation provides to our knowledge the first
ab initio theoretical prediction of the dielectric constant for these 
materials.

In the present case, only a few  independent elements
of the tensors described above are needed, namely
those containing 
derivatives of the total energy and polarization with respect to 
the lattice constant $c$ and the internal structure parameter $u$.
The piezoelectric constant and the Born effective charge
involved are
\be \nonumber
e_{33} = c_0 \left.\frac{\partial P_3}{\partial c}\right\vert_u 
            + 2 e Z^*_{33} \Phi^{-1}_{33} \Xi_{33}\, , 
\ \ \ \
\left.
Z^*_{33} = \frac{\sqrt{3} a^2_0}{4e} \frac{\partial P_3}{\partial u}
				       \right\vert_c.
\ee
The force constant $\Phi_{33}$ (whence
$\Phi_{33}^{-1}=1/\Phi_{33}$ is obtained)
and the 
 internal strain parameter $\Xi_{33}$ are calculated 
as 
derivatives of the Hellmann-Feynman force $F_{3}$ with respect to an
atomic  displacement from equilibrium, and
 to a 
homogeneous strain of the lattice structure, respectively:
\be \nonumber
\left.
\Phi_{33}  =    c^{-1}_0  \frac{\partial F_3}{\partial u}
					   \right\vert_c ,
\ \ \ \ \left.
\Xi_{33} = \frac{4}{\sqrt{3} a^2_0} 
	   \frac{\partial F_3}{\partial c}\right\vert_u.
\ee
The relevant  inverse elastic constant is 
$\lambda^{-1}_{33} = 1/ \lambda_{33}$, where
\be \nonumber
\lambda_{33} = c_0 \frac{\partial \sigma_3}{\partial c}
	    - V \Xi_{33} \Phi^{-1}_{33} \Xi_{33}.
\ee

All calculations  are done using density 
functional theory in the local density approximation (LDA) to describe
the exchange and correlation energy, and ultrasoft 
pseudopotentials~\cite{USPP} for the electron-ion interaction. 
A plane-wave basis cut off at 25 Ry and 12-point  
Chadi-Cohen~\cite{CC} mesh are
found to give fully converged values for the bulk properties.
Given their known importance
~\cite{Fiorentini.PRB}, the semicore {\it d}  states
of Ga and In are  included in the valence.
The piezoelectric tensor and Born charges have been calculated 
\cite{piezo2} via the Berry phase technique \cite{polariz,nota} using
a 16-point Monkhorst-Pack~\cite{MP} k-point mesh in the $a$-plane
direction and 10 point uniform mesh in the $c$ direction,
testing convergence up to 360 total points.
For the  $\varepsilon^\infty$ supercell calculation,
we have employed 
(0001)-oriented superlattices of typically 4 formula units (16 atoms),
and  typical  cation 
sublattice displacements along (0001) of 1-2 \% of the bond
length  for AlN and GaN.
 Smaller displacements ($\sim 0.3$ \%) were used for InN.
25 Ry cutoff and 
12 k-points in the irreducible superlattice Brillouin zone
guarantee convergence. 
No  ionic relaxation is allowed, so that the response is purely
electronic.

We report in Table \ref{tab.eps}  the calculated  dielectric
constants 
and their various components as given by Eq. (\ref{eq.e0})
for AlN, GaN, and InN, along with the available experimental 
data \cite{ref.a,ref.b,ref.c}.
The electronic dielectric constants are very close to the experimental
values for both AlN and GaN, slightly larger for InN. 
The calculated  static constant agrees well with
 experiment for  GaN, the only one for which it 
is available experimentally. We also list
 the values of the  various constants contributing to
$\varepsilon^0$  in Table \ref{tab.s}.
For AlN and GaN, the structural constants are 
 similar, while  the piezoelectric coefficient $e_{33}$ is much higher
 in AlN. InN behaves somewhat differently, as expected from previous
experience on other In compounds \cite{DFPT}.

\paragraph*{Discussion -- }
In homopolar semiconductors,  the external field does  not
cause distortions of the crystal lattice, so that the
static dielectric constant coincides with the electronic one;
our method as outlined above is not applicable in this case,
since no zero-field polarization can exist
in these materials. It does apply, however, to all heteropolar materials,
in which a polarization can always be induced by appropriate atomic 
displacements.
Apart from the
electronic response, a lower  crystal symmetry such as in wurtzites
enables the action of  screening mechanisms related to  lattice
distortions. As apparent from of Eq. (\ref{eq.e0}), the difference among
$\ez$ and $\einf$ is due to  the polarization induced by
optically-active lattice  vibrations 
 (as quantified by the dynamical charges), and to the piezoelectric 
response, if any, along the appropriate axis (as measured by the
piezoelectric constants). 
In a previous work \cite{piezo2} we have
shown that AlN, GaN, and InN have large effective charges, and the highest 
piezoelectric coefficients amongst all tetrahedrally bonded semiconductors. 
Indeed (see Table \ref{tab.eps}), the difference between static and
high-frequency dielectric constants  is large: in III-V nitrides the 
phonon-related term  is comparable to the electronic one, and the
piezoelectric component is about  10\% of each of the other two. 
An important point to be noted  is that the piezoelectric 
contribution cannot be neglected in an accurate calculation. This will
be even more  important in materials with large  piezoelectric
constants, such as ferroelectric perovskites.

\par 
In conclusion, we have presented  a novel procedure for the 
calculation of the dielectric tensor based on the 
geometric quantum phase polarization theory.
The method only uses bulk calculation, with the exception of
 a small supercell calculation, needed in
the determination of the electronic dielectric constant.
As an application, we have provided the first ab initio prediction of the
dielectric  constants of wurtzite AlN, GaN,  and InN.

V.F and F.B. acknowledge partial support by the European Community
through Contract BRE2-CT93, and by CINECA Supercomputing Center, Bologna,
Italy,  through Supercomputing  Grants. D.V. acknowledges support of 
ONR Grant N00014-97-1-0048.


\narrowtext
\begin{table}[h]
\caption{Calculated high-frequency and static dielectric constant 
in AlN, GaN, and InN compared with available experimental data (in
parenthesis).
 $\vare^a$ and $\vare^b$ are the first and second 
 terms in Eq.~(\protect\ref{eq.e0}) respectively, and 
$\ez_{33} = \vare^a + \vare^b + \einf_{33}$.}
\begin{tabular}{lcccccc}
&$\vare^a$&$\vare^b$  &\mcol{2}{c}{$\einf_{33}$}& \mcol{2}{c}{$\ez_{33}$} \\
 \hline
AlN   & 5.06  & 0.64 & 4.61 & (4.68$^a$)    &  10.31 & --- \\

GaN   & 4.44  & 0.15 & 5.69 & (5.70$^b$)    &  10.28 & (10.4$^c$) \\

InN   & 5.51  & 0.61 & 8.49 & (8.40$^b$)    &  14.61 & 
\end{tabular}
$^a$From Ref.~\onlinecite{ref.a}, 
$^b$Ref.~\onlinecite{ref.b} and 
$^c$Ref.~\onlinecite{ref.c}. 
\label{tab.eps}
\end{table}

\vbox{\begin{table}[h]
\caption{ Constants needed in the evaluation of  $\varepsilon^0$
in  AlN, GaN, and InN (see text).}
\begin{tabular}{lccccc}
      &$e_{33}$&$Z^*_{33}$&$\Phi_{33}$ & $\Xi_{33}$ &
      $\lambda_{33}$  \\ \hline
units & C/m$^2$ & --- & N/m& 10$^{21}$ N/m$^3$ & $10^{11}$N/m$^2$  \\
\hline
AlN   & +1.462  & --2.70 & +204     & +0.45  & +3.81 \\

GaN   & +0.727  & --2.72 & +209     & +0.38  & +3.80 \\
 
InN   & +1.092  & --3.02 & +155     & +0.32  & +2.22      
\end{tabular}
\label{tab.s}
\end{table}}
\end{multicols}

\begin{thebibliography}{999}
\bibitem{nota2}
Strictly speaking $\ez_{33}\vert_a$ is 
 the dielectric constant for a material matched to a substrate of
 lattice  constant $a$. The appropriate constant
for a free-standing material would be 
 $\ez_{33}$, since the piezoelectric
 response affects $a$ as well as $c$. The situation envisaged here is
however the one most likely to be realized in practice.
\end{thebibliography}
\end{document}